\documentclass[journal=jacsat,manuscript=article]{achemso}

\usepackage[version=3]{mhchem} 
\usepackage{wasysym}
\usepackage{multirow}
\usepackage{amssymb}
\usepackage{upgreek}
\usepackage{ascii}
\usepackage{array}
\newcolumntype{C}[1]{>{\centering\let\newline\\\arraybackslash\hspace{0pt}}m{#1}}

\author{Pardis Sahafi}
\affiliation{Department of Physics, University of Waterloo, Waterloo, ON, Canada, N2L3G1}
\alsoaffiliation{Institute for Quantum Computing, University of Waterloo, Waterloo, ON, Canada, N2L3G1}
\author{William Rose}
\affiliation{Department of Physics, University of Illinois at Urbana-Champaign, Urbana, Illinois 61801, USA}
\author{Andrew Jordan}
\affiliation{Department of Physics, University of Waterloo, Waterloo, ON, Canada, N2L3G1}
\alsoaffiliation{Institute for Quantum Computing, University of Waterloo, Waterloo, ON, Canada, N2L3G1}
\author{Ben Yager}
\affiliation{Department of Physics, University of Waterloo, Waterloo, ON, Canada, N2L3G1}
\alsoaffiliation{Institute for Quantum Computing, University of Waterloo, Waterloo, ON, Canada, N2L3G1}
\author{Mich\`{e}le Piscitelli}
\affiliation{Department of Physics, University of Waterloo, Waterloo, ON, Canada, N2L3G1}
\alsoaffiliation{Institute for Quantum Computing, University of Waterloo, Waterloo, ON, Canada, N2L3G1}
\author{Raffi Budakian}
\email{rbudakian@uwaterloo.ca}
\affiliation{Department of Physics, University of Waterloo, Waterloo, ON, Canada, N2L3G1}
\alsoaffiliation{Institute for Quantum Computing, University of Waterloo, Waterloo, ON, Canada, N2L3G1}
\alsoaffiliation{Canadian Institute for Advanced Research, Toronto, ON, Canada, M5G1Z8}

\title{Ultra-low dissipation patterned silicon nanowire arrays for scanning probe microscopy}

\date{\today}
  
\begin{document}
\begin{abstract}
In recent years, self-assembled semiconductor nanowires have been successfully used as ultra-sensitive cantilevers in a number of unique scanning probe microscopy (SPM) settings. We describe the fabrication of ultra-low dissipation patterned silicon nanowire (SiNW) arrays optimized for scanning probe applications. Our fabrication process produces, with high yield, ultra-high aspect ratio vertical SiNWs that exhibit exceptional force sensitivity. The highest sensitivity SiNWs have thermomechanical-noise limited force sensitivity of ${9.7\pm0.4\,\text{aN}/\sqrt{\text{Hz}}}$ at room temperature and ${500\pm20\,\text{zN}/\sqrt{\text{Hz}}}$ at 4~K. To facilitate their use in SPM, the SiNWs are patterned within $7\,\upmu\text{m}$ from the edge of the substrate, allowing convenient optical access for displacement detection.
\end{abstract}

Atomic force microscopy (AFM) is a highly versatile technique for probing interactions taking place near surfaces \cite{Binnig_1986,Giessibl_2006}. At the heart of AFM is the mechanical sensor, or cantilever, that can be functionalized to detect a variety of forces that arise from electrical \cite{Girard_2001, Nakamura_2007}, magnetic, \cite{Rugar_1990, Grutter_1992} and chemical \cite{Noy_1995, Lantz_2001, Ducker_1991} interactions. The mechanical properties of the cantilever, e.g., its dimensions, material composition, spring constant, and flexural frequency can all be engineered to provide optimum performance in a particular measurement setting. For applications that require the highest force sensitivity, such as ultra-sensitive spin\cite{Nichol_2012, Rugar_2004, Degen_2009, Poggio_2010} and mass\cite{Vidal_lvarez_2015, Wang_2010, Yang_2006} detection, it is essential to minimize the mechanical dissipation. The demand for ever higher force sensitivity has created a trend towards miniaturization of the cantilever. Aided by advances in self-assembly methods, ultra-sensitive mechanical sensors are being created from a diverse class of materials, including 2D materials, e.g., graphene \cite{Bunch_2007}, and MoSe2\cite{Morell_2019}, semiconductor nanowires (NWs) \cite{Poggio_2019}, and carbon nanotubes\cite{Moser_2013}. Their unprecedented force sensitivity opens new avenues of exploration in metrology, quantum sensing, and fundemental scientific research.

SPM requires locating the force sensor in close proximity to a surface while detecting its motion with high precision, constraining its shape, size, and form factor. In many respects, NWs make ideal nanomechanical sensors. They can be grown vertically with one free end for approaching surfaces and have low intrinsic mechanical dissipation. In recent years, free-standing NW cantilevers have been successfully utilized in several unique SPM settings, including ultra-sensitive force-detected magnetic resonance imaging of nuclear spins\cite{Nichol_2012, Nichol_2013, Rose_2018}, detection of electrical\cite{Rossi_2016}, magnetic\cite{Rossi_2019}, and optomechanical forces\cite{Tavernarakis_2018, Gloppe_2014}, as well as for vectorial force sensing\cite{Gloppe_2014, Rossi_2016, Lepinay_2017}. While these results highlight the versatility of NW sensors, new fabrication techniques are needed to improve and expand their SPM applications.

\begin{figure}[!htb]
        \center{\includegraphics[width=\textwidth]
        {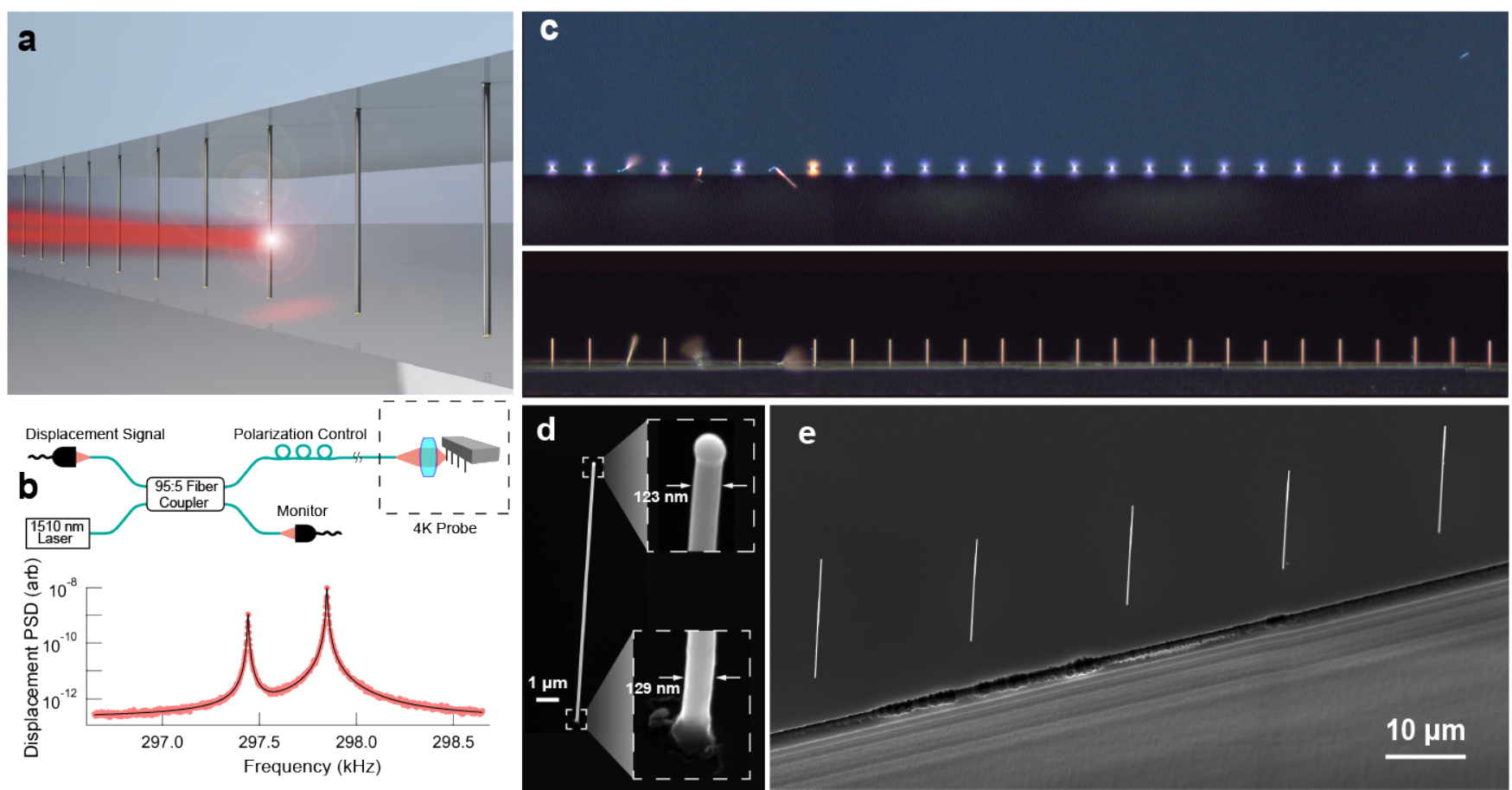}} 
        \caption{\label{Fig_1V2} (a) Rendering of SiNW array with displacement detection laser. (b) Top: schematic diagram of the interferometer. Bottom: characteristic power spectral density (PSD) of thermal oscillations of the two fundamental NW flexural modes. (c) Optical image of SiNWs, from array 1, grown on the sample edge from above (top photo) and from the side (bottom photo).  Each dot in the top photo represents a vertical SiNW. (d) SEM of a single SiNW with insets showing the diameter at the tip and base. (e) SEM of vertical SiNWs at the sample edge.}
\end{figure}

In this work we describe the fabrication of vertical SiNWs optimized for SPM. Although the synthesis of SiNWs is a very mature field, previous work has not focused on their use as SPM force sensors. We present a robust fabrication process that produces arrays of ultra-low dissipation NWs with high yield. Arrays with two different diameter SiNW were fabricated for this work: Array~1 (132~nm diameter, 23~$\upmu\text{m}$ long) and Array~2 (77~nm diameter, 23~$\upmu\text{m}$ long). We characterized the mechanical properties of a large number of SiNWs for each array as a function of temperature from 295~K to 4~K and found that the frequency, spring constant, and quality factor of individual SiNWs were highly consistent within a given array. While all of the SiNWs measured for this work exhibited extremely high force sensitivity, to our knowledge, the 77~nm diameter SiNWs consistently had among the highest force sensitivity measured  in published work, for the range of temperatures that were studied---$9.7\pm0.4\,\text{aN}/\sqrt{\text{Hz}}$ at 295~K, $3.4\pm0.2\,\text{aN}/\sqrt{\text{Hz}}$ at 77~K and $500\pm20\,\text{zN}/\sqrt{\text{Hz}}$ at 4~K. We further present a study of displacement sensitivity achieved using optical interferometry, and discuss the relationship between displacement sensitivity and optical heating effects measured at 4~K. We believe that the fabrication techniques and results presented in this work will facilitate the application of SiNW mechanical sensors by the wider SPM community for a host of novel applications.

To achieve high force sensitivity, it is essential to minimize the thermal fluctuations experienced by the cantilever. The spectral density of force fluctuations for a cantilever with resonance frequency $\omega$, spring constant $k$ and quality factor $Q$ is $S_F=4k_BT\Gamma,$ where $T$ is the cantilever temperature, $k_B$ is the Boltzmann constant, and $\Gamma=k/(\omega Q)$ is the mechanical dissipation. Minimizing $\Gamma$ requires fabrication of cantilevers for which the ratio $k/\omega$ is made as small as possible, while maintaining a high $Q$. An ideal singly-clamped cylindrical beam has two fundamental orthogonal flexural modes of degenerate frequency. For these two modes, $k$ and $\omega$ can be expressed in terms of the radius $R$ and length $L$ of the beam using Euler-Bernoulli beam theory.
\begin{align}
    k&=\left(\frac{3\pi}{4}\right)\left(\frac{R^4}{L^3}\right)E,\label{Eq_k}\\
    \omega&=\left(\frac{1.875^2}{2}\right)\left(\frac{R}{L^2}\right)\sqrt{\frac{E}{\rho}}\label{Eq_omega},
\end{align}
where $E$ is the Young's modulus, and $\rho$ is the mass density. The dissipation can then be expressed as
\begin{equation}\label{Eq_Gamma2}
    \Gamma = 1.841\left(\frac{R^3}{LQ}\right)\sqrt{E\rho}.
\end{equation}
Eq.~(\ref{Eq_Gamma2}) shows that mechanical dissipation, and therefore the thermomechanical force noise, is minimized by fabricating long, thin beams with high quality factors from materials for which the product $E\rho$ is small.

Within the family of semiconducting NWs, silicon offers a number of key advantages that make it an outstanding material for fabricating force sensors. It has a relatively low density and NWs can be grown vertically with nearly zero taper, enabling the fabrication of ultra-high aspect ratio cylindrical structures. Such nanostructures are capable of withstanding large strain without breaking. Silicon processing is well-developed and there are well-established methods of functionalizing the surface of silicon tailored for a wide variety of applications.

From the standpoint of optimizing force sensitivity, it seems obvious that we should opt for fabricating the thinnest and longest NWs possible. However, our choice of SiNW dimensions are guided by two other important considerations.  First, we require the minimum SiNW diameter to be no smaller than about 50~nm, because below this diameter it becomes significantly more challenging to achieve high displacement sensitivity using optical techniques. We address this point in detail later in the paper. Second, it has been observed that as the cantilever is brought close to a surface, force fluctuations caused by short-range tip-surface interactions, can degrade mechanical performance\cite{Stipe_2001_M, Kuehn_2006, Volokitin_2005}. Of particular relevance to NW sensors are low frequency charge fluctuations near the NW-surface interface, which produce fluctuating force gradients that give rise to fluctuations in the spring constant of the NW. A shift $\delta k$ in the spring constant will produce a change in frequency $\delta\omega=\omega\delta k/(2k)\propto \delta k(L^3/R^4)$. Hence, as the aspect ratio $L/R$ is made larger, the NW sensor will suffer increased frequency jitter, which limits its usefulness in force sensing. Our choice of dimensions was informed by previous MRFM measurements, where SiNWs with diameter between 50--100~nm and length up to 20~$\upmu\text{m}$ were be used with minimal frequency jitter within 50~nm of the surface.

The SiNW arrays were grown epitaxially using the vapor-liquid-solid method \cite{Wagner_1964} on a Si[111] substrate from a patterned array of Au catalyst particles. Growth occurred at 550~$^\circ\text{C}$ in a $\text{H}_2$ and HCl atmosphere using a $\text{SiH}_4$ precursor. The $\text{H}_2$ carrier gas terminates the Si bonds, modifying the Au wetting parameters and inhibiting Au migration, while the addition of HCl minimizes the breakup of the Au catalyst and also etches the sidewalls of the SiNW during growth. These factors promote the growth of long vertical SiNWs with very low taper and smooth side walls\cite{Gentile_2012,Oehler_2009}.

Two arrays with different diameter Au nanodisks were patterned by e\nobreakdash-beam lithography. Array~1 was fabricated by exposing 100~nm diameter circular areas, separated by 20~$\upmu\text{m}$, in a MMA/PMMA bilayer resist. After development, 52~nm of Au was evaporated, and the excess Au along with the e\nobreakdash-beam resists are removed by lift-off. Array~2 was fabricated by exposing 50~nm diameter disks separated by 15~$\upmu\text{m}$ in a ZEP resist layer, and depositing 33~nm of Au. Immediately after lift-off, the samples underwent rapid thermal annealing in an Ar atmosphere. The annealing procedure leads to the formation of a Au/Si eutectic \cite{Pinardi_2009, Ressel_2003}, ensuring that the Au-Si interface remains stable throughout the remaining process steps. We found that the annealing procedure was critical to having a high yield of vertical SiNWs. The silicon substrate was then etched by deep reactive ion etching into $1.0\times1.5$~mm pieces, leaving a single row of nanodisks, approximately 7~$\upmu\text{m}$ from the 1~mm wide edge of each piece. The close proximity of the SiNWs to the edge of the substrate allows optical access to each SiNW for displacement detection. Fig.~\ref{Fig_1V2} shows a schematic, along with optical and SEM images, of the fabricated SiNW devices. Further details regarding the fabrication procedure are provided in the Supporting Information section. 

Table~\ref{Tab_nwdims} provides a summary of the dimensions of the SiNWs, along with the mean calculated fundamental frequency and spring constant for each array. These values were calculated using Eqs.~(\ref{Eq_k}, \ref{Eq_omega}), with $\rho=2.33$ g/cm$^3$, $E=169$ GPa \cite{McSkimin_1964}, and the measured dimensions for (33) 132~nm diameter and (18) 77~nm diameter SiNWs.  As the taper was sufficiently small for all the SiNWs, we assumed a single value for the diameter, given by the average of the base and tip diameters. Array~1 had a very high yield of 33/50 high quality, vertical SiNWs. Array~2 was further optimized for high force sensitivity by significantly reducing the radius to 77~nm, which resulted in a somewhat lower yield of completely vertical SiNWs. However, the vertical SiNWs from both sets were very uniform, with small standard deviations in diameter and length, and had very little tapering from base to tip.

\begin{table}
\begin{tabular}{C{15mm}	C{20mm}	C{15mm} C{20mm} C{20mm} C{20mm}}
\hline \hline \\[-12pt]
		& $\diameter$	& $\Delta\diameter$	& L		& $k$			& $f_{\text{calc}}$		\\
		& (nm)		& (nm)			& ($\upmu$m)		& ($\upmu$N/m)	& (kHz)			\\ \hline \\[-10pt]
Array 1	& 132 $\pm$ 10	& 4 $\pm$ 4			& 23 $\pm$ 1.8	& 610 $\pm$ 176	& 291 $\pm$ 49		\\[2pt]
Array 2	& 77 $\pm$ 1	& 1 $\pm$ 1			& 23 $\pm$ 0.3	& 68 $\pm$ 5	& 168 $\pm$ 5		\\[1pt]
\hline \hline
\end{tabular}

\caption{Mechanical properties of SiNW arrays. From left to right, the table lists the average diameter $(\diameter)$, difference in diameter at the base and tip $(\Delta\diameter)$, length (L), and the calculated spring constant $(k)$ and fundamental flexural frequency $(f_{\text{calc}})$  for the $\diameter$132~nm  and $\diameter$77~nm SiNW arrays. The measured values were obtained from SEM images of each SiNW. Mean values and standard deviations were calculated for 33 $\diameter$132~nm and 18 $\diameter$77~nm SiNWs.}\label{Tab_nwdims}
\end{table}

To characterize the mechanical properties of each SiNW, we measured the frequencies and quality factors of their two fundamental flexural modes at 295~K, 77~K and 4~K in high vacuum using the polarized interferometry setup shown in Fig.~\ref{Fig_1V2}b. Due to small shape asymmetry, the mode frequencies were slightly split and distinguishable. The interferometer provided excellent sensitivity for measuring SiNW displacement along the axis of optical propagation \cite{Nichol_2008}. A typical thermal displacement spectrum is shown in Fig.~\ref{Fig_1V2}b. The frequency and quality factor for the two lowest frequency flexural modes were determined by fitting the double-resonance lineshape, given by Eq. (\ref{S_omega}),  to the thermal displacement power spectral density.
\begin{equation} \label{S_omega}
S(\omega)=\frac{\alpha_1^2}{(\omega^2-\omega_1^2)^2+\omega^2\omega_1^2/Q_1^2} + \frac{\alpha_2^2}{(\omega^2-\omega_2^2)^2+\omega^2\omega_2^2/Q_2^2}
\end{equation}
In equation (\ref{S_omega}), $\alpha_{1,2}$ are the relative amplitudes, $\omega_{1,2}$ are the resonant frequencies and $Q_{1,2}$ are the quality factors of the two fundamental flexural modes. The ratio $\alpha_1/\alpha_2$ is determined by the orientation of each mode with respect to the optical axis of the interferometer.
The results of the measurements are summarized in Table \ref{Tab_nwQs}.

\begin{table}
\caption{Mean and standard deviation of frequency, frequency splitting, quality factor, mechanical dissipation, and force noise spectral density at different bath temperatures for the $\diameter$132~nm (top row) and $\diameter$77~nm (bottom row) SiNWs.}
\label{Tab_nwQs}
\begin{tabular}{C{15mm}	C{20mm}	C{20mm}	C{20mm}	c 	c 	}
\hline \hline \\[-12pt]
$T$			& $f_{\text{meas}}$	& $\Delta f/f$	& $Q$			& $\Gamma$				& $S_F^{1/2}$			\\
(K)			& (kHz)			& ($10^{-3}$)	& ($10^3$)		& ($10^{-15}$ kg s$^{-1}$)	& ($10^{-18}$ N~$\text{Hz}^{-1/2}$)	\\[1pt] \hline \hline \\[-12pt]
\multirow{2}*{295}	& 302 $\pm$ 54		& 2 $\pm$ 1		& 12 $\pm$ 1	& 26 $\pm$ 4			& 20 $\pm$ 2			\\
			& 170 $\pm$ 6		& 6 $\pm$ 12	& 11 $\pm$ 1	& 5.7 $\pm$ 5			& 9.7 $\pm$ 0.4			\\ \hline \\[-13pt]
\multirow{2}*{77}   & 305 $\pm$ 54		& 2 $\pm$ 1		& 23 $\pm$ 2	& 8.4 $\pm$ 1.3			& 6.0 $\pm$ 0.4			\\
			& 171 $\pm$ 6		& 6 $\pm$ 12	& 23 $\pm$ 2	& 2.7 $\pm$ 0.3			& 3.4 $\pm$ 0.2			\\ \hline \\[-13pt]
\multirow{2}*{4}	& 305 $\pm$ 55		& 2 $\pm$ 1		& 55 $\pm$ 5	& 5.8 $\pm$ 1			& 1.2 $\pm$ 0.1			\\
			& 171 $\pm$ 6		& 6 $\pm$ 12	& 59 $\pm$ 4	& \textbf{1.1} $\pm$ 0.1		& \textbf{0.50} $\pm$ 0.02	\\
\hline \hline
\end{tabular}
\end{table}

Fig. \ref{Tab_nwQs} shows the distribution of force noise spectral density measured for all of the the 132~nm and 77~nm diameter SiNWs studied. Our fabrication process produces vertical SiNWs with highly consistent mechanical properties, extremely low mechanical dissipation, and ultra-high force sensitivity.  For comparison, a single electron spin placed in a magnetic field gradient of $1\times10^6$~T/m will produce a peak force of 9.3~aN, which is approximately equal to the 9.7~aN~RMS force noise in the 1~Hz bandwidth of the 77~nm diameter SiNWs at room temperature. The same diameter SiNWs would be capable of detecting 0.5~aN~RMS in a 1~Hz bandwidth at 4~K, equal to just 51 proton spins in a peak field gradient of $1\times10^6$~T/m. For reference, static magnetic field gradients as large as $6\times10^6$~T/m have been achieved using a nanoscale dysprosium sources in magnetic resonance force microscopy experiments. \cite{Mamin_2012} The combination of ultra-high force sensitivity and high magnetic field gradients would greatly advance the goal of atomic scale magnetic resonance imaging. In addition, the ability to fabricate patterned arrays of SiNWs having consistently high force sensitivity would be useful to a host of other ultra-sensitive SPM applications.

\begin{figure}[!htb]
        \center{\includegraphics[width=120mm]
        {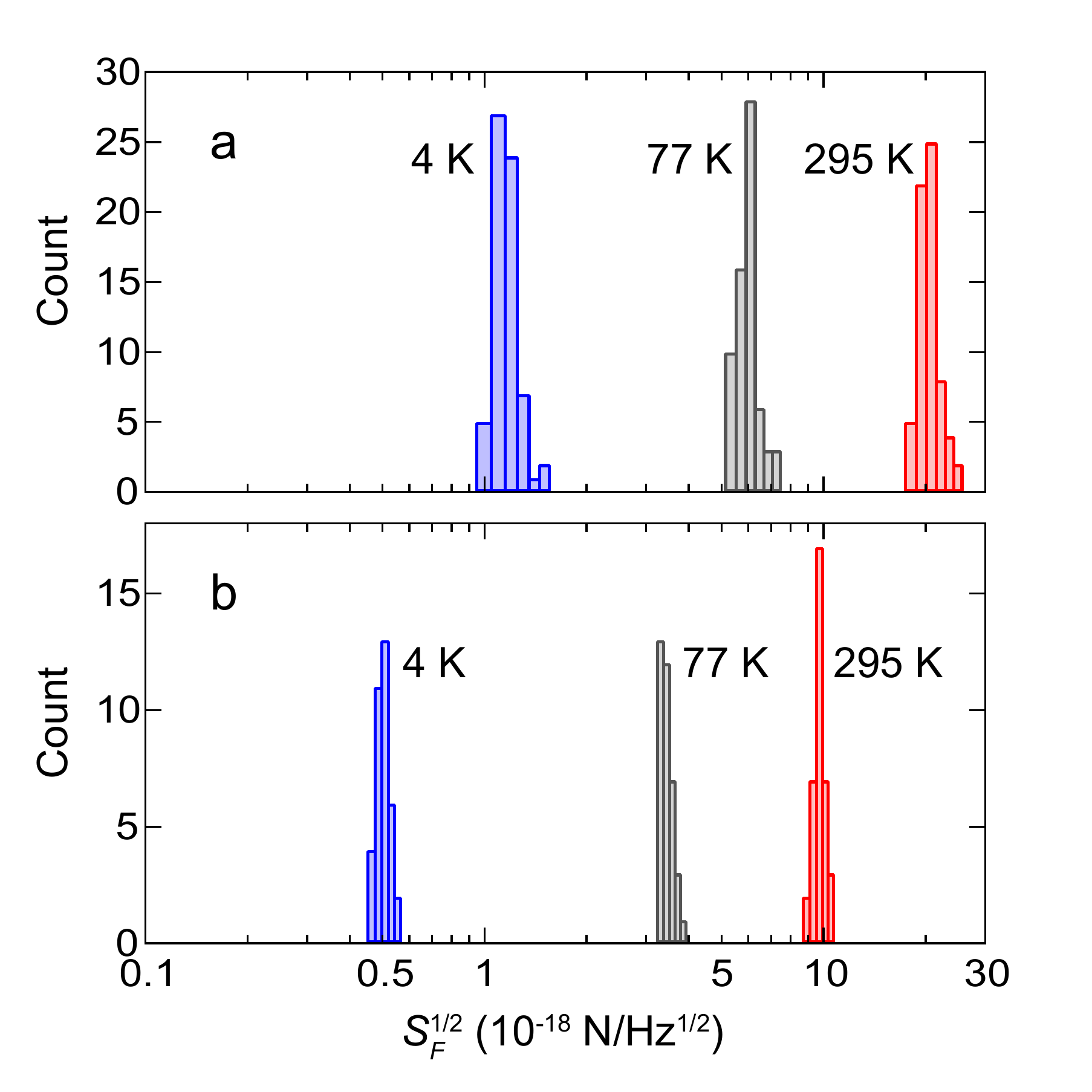}}
        \caption{\label{Fig_Sf} Distribution of force noise spectral density obtained for the two lowest frequency flexural modes for the $\diameter$132~nm (a) and $\diameter$77~nm (b) arrays. Data were obtained for (33) SiNWs for (a) and (18) SiNWs for (b).}
\end{figure}

Detecting displacement with high sensitivity is essential for the implementation of NW force sensors in SPM. Laser interferometry is the most versatile, high-sensitivity, non-invasive method for displacement detection. However, as the diameter of the NW becomes smaller than a few hundred nanometers, the reflected light from the NW decreases significantly, requiring higher incident light levels to be used. For low temperature applications, it is very important to consider heating caused by optical absorption. Previous work has shown that the thermal conductivity of SiNWs are reduced significantly relative to the bulk value, decreasing with NW diameter\cite{Li_2003}. In this confined geometry thermal boundary scattering dominates over phonon-phonon scattering shifting the Umklapp peak to higher temperatures. The low thermal conductivity and high aspect ratio produces a large thermal resistance, such that picowatts of absorbed power can lead to several degrees of heating at 4~K.

To characterize optical heating effects, we measured the displacement sensitivity and the SiNW temperature as a function of the incident optical power. The measurements were conducted at a base temperature of 4.2~K using an optical fiber-based interferometer operating at 1510~nm. The light exiting the fiber was focused to a 2~$\upmu\text{m}$ diameter spot near the tip of the SiNW. The SiNW temperature was determined by integrating the thermal displacement power spectral density, and assuming equipartition of thermal energy. Further details regarding the determination of the SiNW temperature are provided in the Supporting Information section.

\begin{figure}[!htb]
        \center{\includegraphics[width=120mm]
        {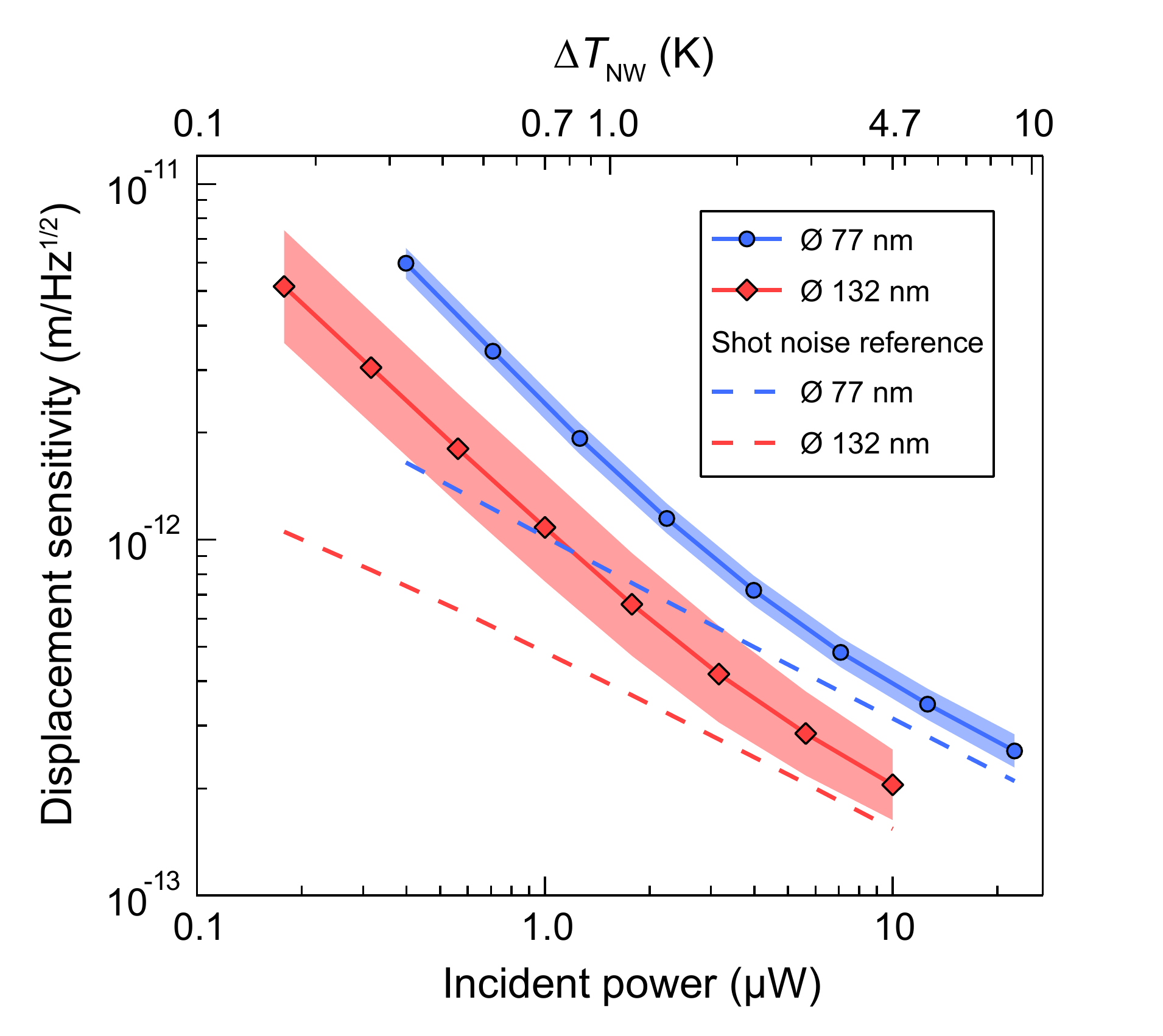}}
        \caption{\label{Fig_dispsens} Displacement sensitivity as a function of incident optical power obtained at 4.2~K for the $\diameter$132~nm and $\diameter$77~nm arrays. The incident power represents the amount of light being focused to a 2~$\upmu$m diameter spot on the SiNW. Data shown are the mean displacement sensitivity of two wires for each array. Shaded areas indicate the range of measured values. The top axis shows heating observed above the base temperature. The dashed lines show the displacement sensitivity if shot noise were the only noise source. For reference, the average thermal displacement of the $\diameter$77~nm SiNWs at 4.2~K is $3.8 \times 10^{-10}$~m/$\sqrt{\text{Hz}}$ within the $\omega/(2\pi Q)=2.9$~Hz bandwidth.}
\end{figure}

The data presented in Fig.~\ref{Fig_dispsens} show the displacement sensitivity as a function of the incident laser power for two representative 132~nm and 77~nm diameter SiNWs. Over the range of our measurement, heating was observed to be insensitive to the diameter of the NWs (see Fig. S7), suggesting that the higher optical absorption of the larger diameter NW is partly compensated by the increase in thermal conductivity. The dashed lines indicate the theoretical limit of displacement sensitivity determined from the photocurrent shot noise. At high light levels the measured displacement sensitivity is close to the theoretical shot-noise limit. However at low light levels, the displacement sensitivity is dominated by the current noise of the amplifier and deviates substantially from the theoretical limit. A discussion of our detection electronics are presented in the Supporting Information section. This result indicates that with lower noise detection electronics it should be possible to operate with substantially lower laser power, significantly reducing optical heating, and possibly extending the range of optical detection techniques into the millikelvin regime.

In summary, we have developed a robust process for fabricating SiNW arrays that exhibit ultra-low mechanical dissipation, with controllable geometry, that are tailored for use in SPM.  The key elements of the fabrication process are (i) the use of HCl, which inhibits the breakup of the Au catalyst particle during growth and promotes the growth of ultra-low taper SiNWs that can be grown with very high aspect ratio, and (ii) maintaining a high quality Au-Si interface throughout the fabrication process, enabling the growth of vertical SiNWs near the edge of the substrate for convenient optical access.  By combining these design elements and further optimizing the growth parameters we were able to fabricate, with high yield, arrays of SiNWs that to our knowledge, exhibit among the lowest dissipation and highest force sensitivity at 4~K of any free-standing nanomechanical force sensors in the published literature. Measurements of optical heating indicate that it is possible to achieve high displacement sensitivity with minimal heating down to 4~K and improvements in the detection electronics or other methods of mitigating optical absorption are needed to extend optical detection techniques to millikelvin temperatures. We believe the results presented in this work will facilitate the use of SiNW mechanical sensors in a host of novel SPM applications.

\section{Acknowledgements}
This work was undertaken thanks in part to funding from the U.S. Army Research Office through Grant No. W911NF1610199, the Canada First Research Excellence Fund (CFREF), and the Natural Sciences and Engineering Research Council of Canada (NSERC). The University of Waterloo's QNFCF facility was used for this work. This infrastructure would not be possible without the significant contributions of CFREF-TQT, CFI, Industry Canada, the Ontario Ministry of Research and Innovation and Mike and Ophelia Lazaridis. Their support is gratefully acknowledged. We are grateful to Vito Logiudice, Nathan Nelson-Fitzpatrick, Greg Holloway, Sandra Gibson and Lino Eugene for useful discussions. We would also like to thank Deler Langenberg for his assistance in the installation of the CVD system.

\bibliography{bibl}

\end{document}


\section{Fabrication and Growth}

We began with a 4~in. diameter intrinsic Si (111) wafer. A 1~$\upmu$m layer of surface oxide was grown in a Tystar Tytan 4600 Mini Fourstack Horizontal Furnace. This oxide is later etched away using a buffered HF solution in order to produce a pristine Si (111) surface. The oxide also protects the sample surface during storage. Sample chips were diced into 15$\times$15~mm chips using a DISCO DAD3240 dicing saw and cleaned of organic and metallic contaminants using an RCA clean consisting of two steps designed to remove organic contaminants (SC\nobreakdash-1) and metallic contaminants (SC\nobreakdash-2). The cleaning solutions used were a 10:2:1 solution of DI Water: 30\% Hydrogen Peroxide (H$_2$O$_2$): 29\% Ammonium Hydroxide (NH$_4$OH) for the SC\nobreakdash-1 process and 10:2:1 solution of DI Water: 30\% Hydrogen Peroxide (H$_2$O$_2$): 30\% Hydrochloric Acid (HCl) for SC\nobreakdash-2. Both solutions were heated to 70~$^\circ$C. Samples were immersed in each solution for 15~min.

Between the two RCA cleaning steps, the sample was immersed in 10:1 buffered oxide etchant for 25~min to remove the 1~$\upmu$m thermal oxide and leave behind a clean, high quality (111) Si surface.

\begin{figure}[ht!]
    \centering
    \includegraphics[scale = 0.65]{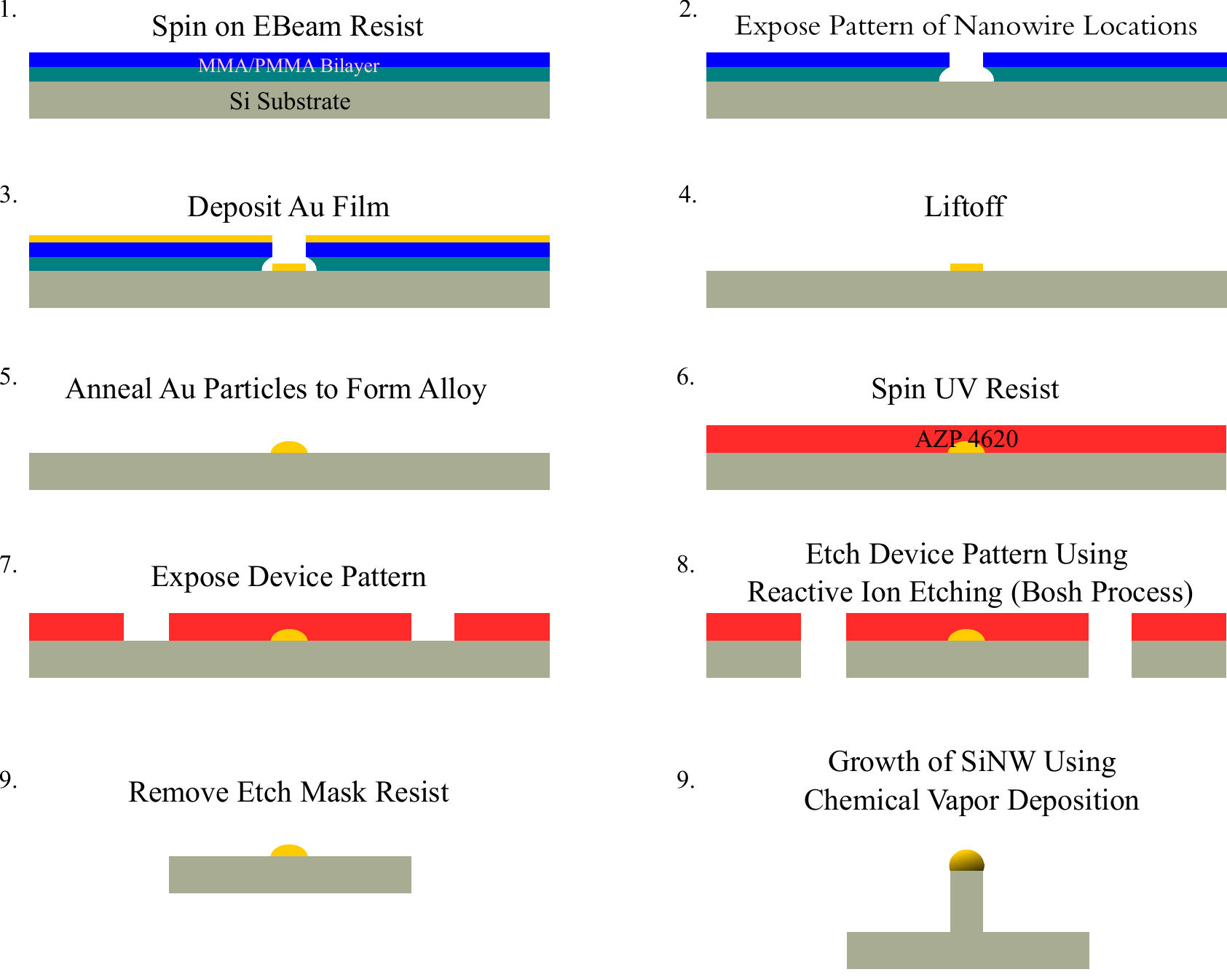}
    \caption{Schematic diagram of the SiNW fabrication process.}
    \label{fig:fab_diagram_1}
\end{figure}

The sample was patterned with gold discs using ebeam lithography in a RAITH150 Two 30kV Direct Write system. For the 130~nm NW arrays, the ebeam resist used was a MMA EL10/PMMA A3 bilayer with a total thickness of approximately 650~nm. This bilayer was chosen in order to produce an undercut surrounding the deposited gold to aid in liftoff after deposition, however this bilayer configuration proved challenging to use for smaller features required to produce 77~nm diameter wires. Arrays of circular holes with a radius of 100~nm were exposed using a 10~kV electron beam delivering an area dose of 600~$\upmu$C/cm$^2$. These samples were developed in a solution of 1:3 Methyl Isobutyl Ketone (MIBK): Isopropyl alcohol for 2~min.  

For 77~nm diameter wires, a 300~nm thick ZEP520-A e\nobreakdash-beam resist was used, with an exposed  hole size of 70~nm. We used a 25~kV beam with an area dose of 520~$\upmu$C/cm$^2$. ZEP resists have a higher dose sensitivity compared to that of MMA/PMMA resists. This allows for an undercut to be formed by back-scattered electrons during the exposure step. These samples were developed in ZED-N50 (n-Amyl Acetate) for 1.5~min.

Prior to deposition, the substrate was immersed in 10:1 buffered oxide etchant for 10~s in order to remove the native surface oxide layer and to protect the surface from oxide formation by terminating dangling bonds on the Si surface \cite{Kang_2002, Bal_2010}, ensuring a good contact between the deposited gold and the substrate. To produce 130~nm wires, a 52~nm film of gold was deposited using an Intlvac Nanochrome II ebeam evaporator at a rate of 1.0~\AA/s at $4\times 10^{-6}$~Torr. Gold liftoff was done by sonication in PG remover heated to 80~$^\circ$C over a period of 30~min. To produce 77~nm diameter wires, 33~nm of gold was deposited at 1.0~\AA/s and lifted off under the same conditions.

\begin{figure}[ht!]
    \centering
    \includegraphics[scale = 0.3]{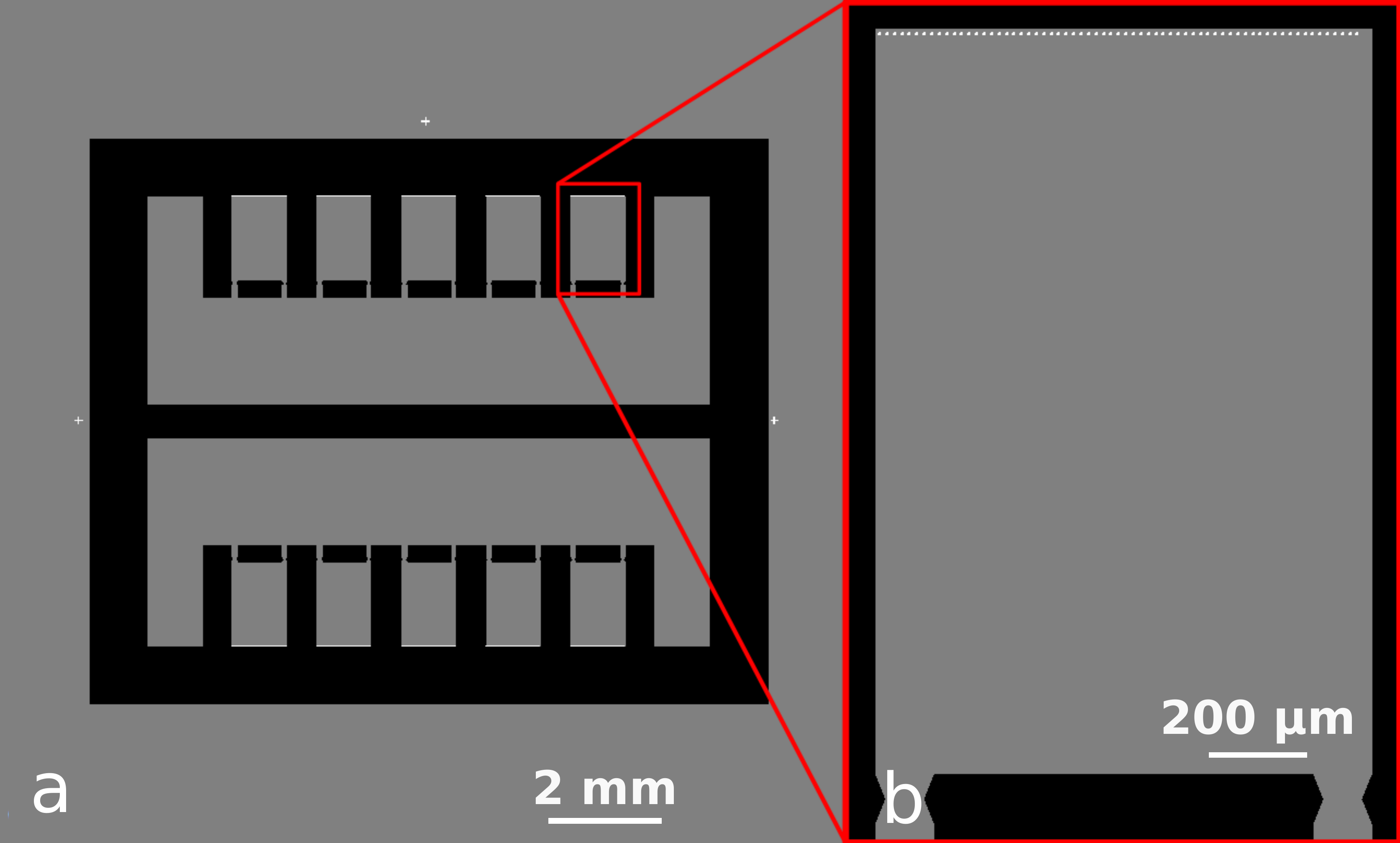}
    \caption{Device schematic. (a) Etch mask pattern on diced 15$\times$15~mm Si piece. Black areas indicate material removed with DRIE. Gray areas are remaining Si. Alignment crosses are used to align the DRIE etch mask pattern to the Au disk arrays. (b) Zoomed in view of an individual 1$\times$1.5~mm Si chip. Tabs at the bottom allow individual chips to be easily broken out. White dots indicate Au disc locations but are not to scale (enlarged for visibility).}
    \label{fig:device_schematic}
\end{figure}

After liftoff, the gold particles were subjected to a rapid thermal anneal in an Allwin AccuThermo AW 610 Rapid Thermal Processor (RTP). Samples were placed in a 5~L/min Ar flow at 400~$^\circ$C for the 130~nm wires and 500~$^\circ$C for the 77~nm wires. This temperature was maintained for 10~min, forming a Au/Si eutectic in order to prevent contamination of the Au-Si interface throughout the remainder of the fabrication procedure. \cite{Pinardi_2009, Ressel_2003}

\begin{figure}[ht!]
    \centering
    \includegraphics[scale = 0.45]{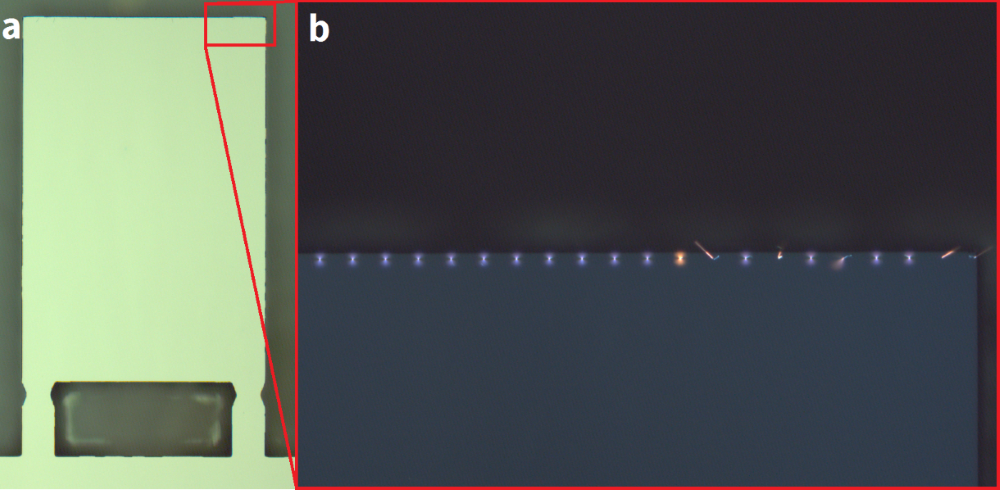}
    \caption{Top down optical microscope images of a single SiNW chip after the growth process. (a) Bright field image showing the etched device shape. Chip dimensions are 1$\times$1.5 mm. (b) Polarized light microscopy image of a chip corner showing a high yield of vertical SiNWs. Vertical NWs are visible as bright spots near the chip edge.}
    \label{fig:optical_iamge_1}
\end{figure}

To define the device pattern, an anisotropic deep reactive ion etch (DRIE) using the Bosch process \cite{Bosch_1996} was used. 700 Bosch cycles were run in an Oxford Instuments PlasmaLab System 100 to completely etch through the substrate to define the final shape of the device. This allowed us to locate NWs at the edge of the chip, allowing for them to be optically accessible. We also found that growing NWs in close proximity to the edge of a chip resulted in a slightly increased yield of vertical NWs compared to NWs patterned in the center if a chip.

The mask (Figure  \ref{fig:device_schematic}) used for the DRIE process was a double layer AZP 4620 photoresist with a total thickness of approximately 21~$\upmu$m. The thickness of the resist allows it to withstand the long etch process, protecting the gold dots during the etch. The device pattern was defined using UV lithography in a Heidelberg MLA 150 Direct Write UV Lithography system and developed in AZ 400K for 4~min. The mask pattern is aligned to the three alignment marks (Figure \ref{fig:device_schematic}) that were patterned during the EBL and Au evaporation steps. Defining the etch mask in this way allows us to align to our patterned gold dots with high accuracy, typically to within 1--2~$\upmu$m of the desired location.

After the DRIE, the etch mask was removed by sonicating in PG remover at 80~$^\circ$C for 40~min. It was found on some samples that localized heating during the etch process led to hardbaking of the etch mask, causing microscopic photoresist contamination to remain even after removal of the mask. This remaining contamination was removed by immersing the sample in the SC-1 solution described above at 70~$^\circ$C for 15~min to remove any remaining organic material. A diagram of the described fabrication process is presented in Figure  \ref{fig:fab_diagram_1}. \vspace{10pt}

\begin{figure}[ht!]
    \centering
    \includegraphics{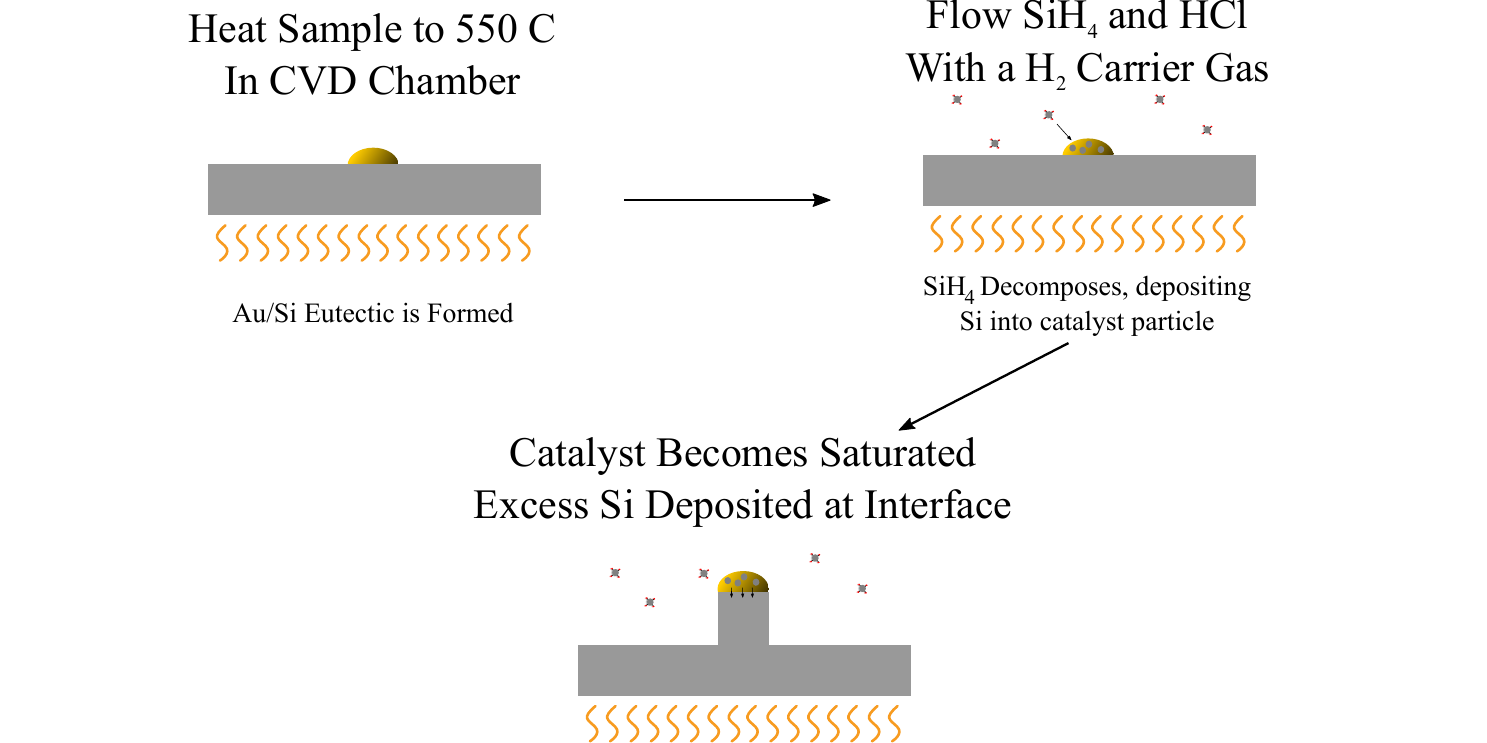}
    \caption{Diagram of SiNW growth via the VLS mechanism.}
    \label{fig:fab_diagram_2}
\end{figure}



Growth of the SiNWs was done in a FirstNano EasyTube 3000EXT Chemical Vapor Deposition (CVD) system using the vapor-liquid-solid (VLS) mechanism \cite{Wagner_1964}. Our growth process takes place at 550~$^\circ$C and an absolute pressure of 4~Torr.

After the final SC\nobreakdash-1 clean, we load our sample into the furnace tube of the CVD system heated to 100~$^\circ$C overnight. This allows us dehydration bake the sample to remove any moisture that may have accumulated during transit to the CVD system. 

The process furnace is ramped to 550~$^\circ$C under a 1~SLM H$_2$ flow. We found that growing at 550~$^\circ$C provided a significantly higher yield of vertical NWs compared to higher temperatures \cite{Schmid_2008}. At temperatures lower than 550~$^\circ$C, we observed a reduced yield of vertical NWs. We also observed that some of the catalyst particles failed to nucleate at lower temperatures, leading to fewer NWs growing overall.

During the temperature ramp, the pressure is kept at 4~Torr by a variable speed pump. Once the furnace temperature has stabilized, process gases are introduced to the system. A 20~SCCM flow of Hydrogen Chloride (HCl) and a 10~SCCM flow of Silane (SiH$_4$) are introduced. These flow rates result in partial pressures of $\mathrm{P_{H_2}} = 3.9$~Torr, $\mathrm{P_{HCl} }= 78$~mTorr, $\mathrm{P}_{\mathrm{SiH_4}}=39$~mTorr. H$_2$ acts as a carrier gas, and allows dangling bonds on the newly grown Si surface to be passivated by hydrogen termination. SiH$_4$ acts as a precursor, providing the source of Si to be deposited via the Au catalyst particle. The presence of HCl has been shown to prevent breakup of the catalyst particle during growth and to suppress radial growth on the NW side walls by etching the side walls during growth. \cite{Oehler_2009,Gentile_2012}. This allows  us to grow NWs with clean side walls, and little to no taper. We observed that the prevention of sidewall growth contributes to achieving a high quality factor. To achieve a SiNW length of 23~$\upmu$m, we maintain the flow of process gases for a period of 3~h. Under these conditions NWs grow at a constant rate, allowing us to control their length by varying the growth time.

Once the growth process is complete, the flow of process gas is cut off and the furnace tube is allowed to cool under a H$_2$ flow to 300~$^\circ$C. The tube is then evacuated, filled with N$_2$, and allowed to cool to room temperature over a period of roughly 1~h under a N$_2$ flow.

\section{Comparison of frequency determination methods}
\begin{figure}[!htb]
        \center{\includegraphics[width=0.9\textwidth]
        {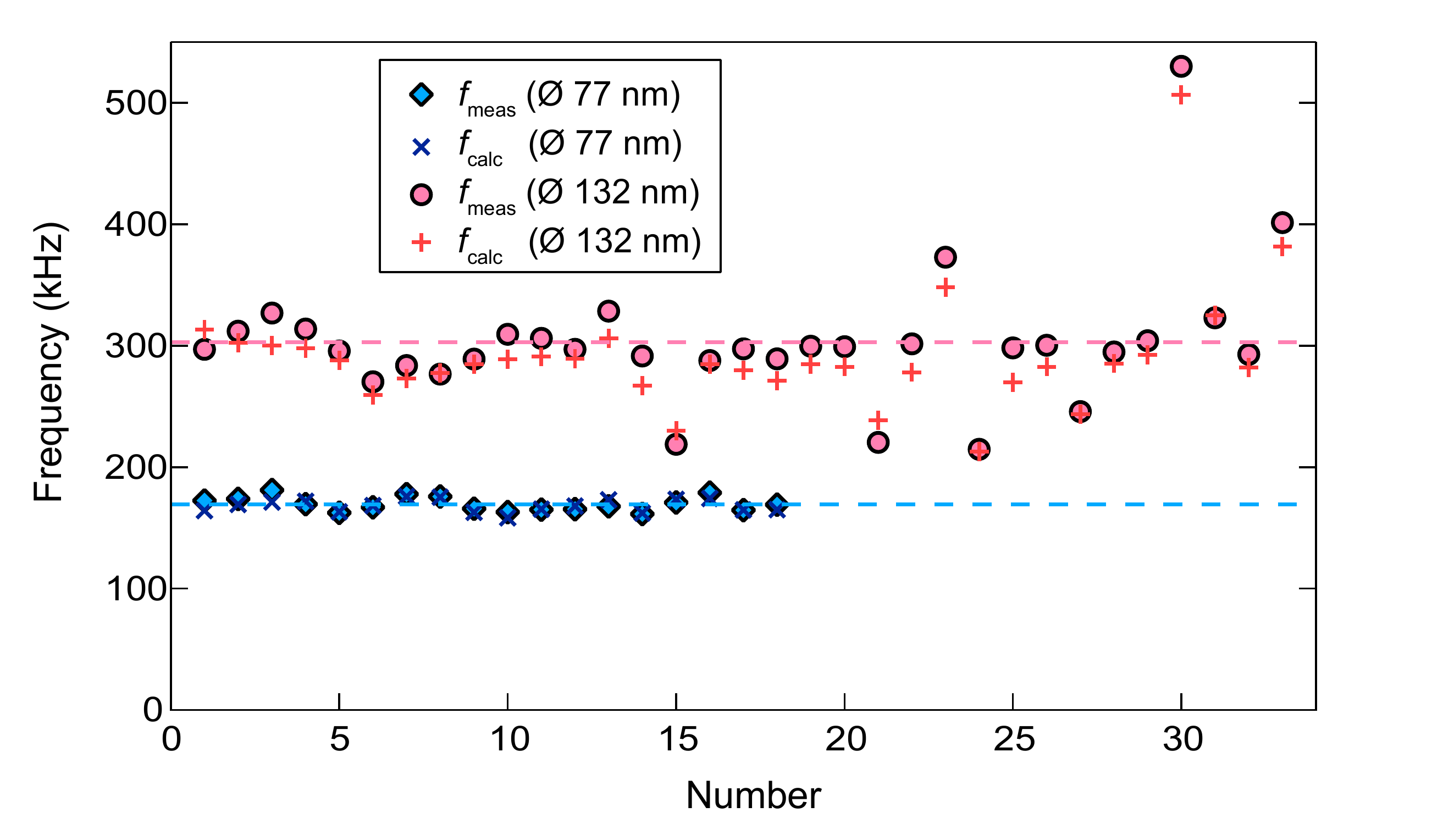}}
        \caption{\label{Fig_fcomp} Measured and calculated frequency values for each SiNW. Measured values displayed are the average of the two lowest frequency modes. Colored dashed lines show the average measured frequency for each array.}
\end{figure}

The mechanical dissipation of a flexural mode of the SiNWs is given by $\Gamma=k/(\omega Q)$, where $k$ is the spring constant, $\omega$ the angular frequency, and $Q$ the quality factor of the mode. The frequencies and quality factors for the two fundamental modes of each wire can be determined accurately by fitting the double-resonance lineshape, Eq. (4), to the power spectral density obtained by interferometry. However, determining $k$ by interferometry is much less straightforward. Instead, $k$'s were calculated by Eq. (1) using radius, $R$, and length, $L$, for each wire as measured by scanning electron microscopy. To confirm the accuracy of these dimension measurements and the validity of this $k$ determination, frequencies measured at room temperature by interferometry were compared to frequencies calculated using Eq. (2), which also depends on $R$ and $L$.  Full data are shown in Fig. \ref{Fig_fcomp} and Table \ref{Tab_fcomp}. Taking $f_{\text{meas}}$ to be the correct value, the average error in $f_{\text{calc}}$ was 3.6\% for the larger SiNWs and 1.2\% for the smaller SiNWs. Because $k$ depends more strongly than $f$ on $R$ and $L$, the error in $k$ may be slightly larger.

\begin{table}
\begin{tabular}{C{14mm} C{18mm} C{18mm} C{18mm} C{12mm}}
\hline \hline \\[-12pt]
		& $\diameter$	& $f_{\text{calc}}$		& $f_{\text{meas}}$	& Error	\\
		& (nm)		& (kHz)		& (kHz)		& (\%)	\\ \hline \\[-9pt]
Array 1	& 132 $\pm$ 10	& 291 $\pm$ 49	& 302 $\pm$ 54	& 3.6		\\[3pt]
Array 2	& 77 $\pm$ 1	& 168 $\pm$ 5	& 170 $\pm$ 6	& 1.2		\\[1pt]
\hline \hline
\end{tabular}
\caption{Mean and standard deviation of SiNW diameter ($\diameter$), resonance frequency calculated from nanowire dimensions using Euler-Bernoulli beam theory ($f_{\text{calc}}$), resonance frequency measured with interferometer ($f_{\text{meas}}$), and error between the two frequency measurement means, calculated by $(f_{\text{meas}}-f_{\text{calc}})/f_{\text{meas}}$.}\label{Tab_fcomp}
\end{table}

\section{Displacement sensitivity and SiNW heating}

A fiber-coupled distributed-feedback 1510 nm laser was used for the interferometer. A splitter sent 95\% of the optical power to a monitor photodiode. The remaining 5\% traveled through a three paddle polarization controller and then into the probe. The fiber end inside the probe was cleaved, which reflected about 3\% light and served as our reference beam. The remaining light was focused by a Lightpath 355631 lens with a working distance of 280~$\upmu$m and a numerical aperture of 0.55. The light was detected with an InGaAs PIN photodiode and amplified with a FEMTO DHPCA-100 transimpedance amplifier. Fig 1b shows a schematic of the interferometer.

\begin{figure}[!htb]
        \center{\includegraphics[width=1\textwidth]
        {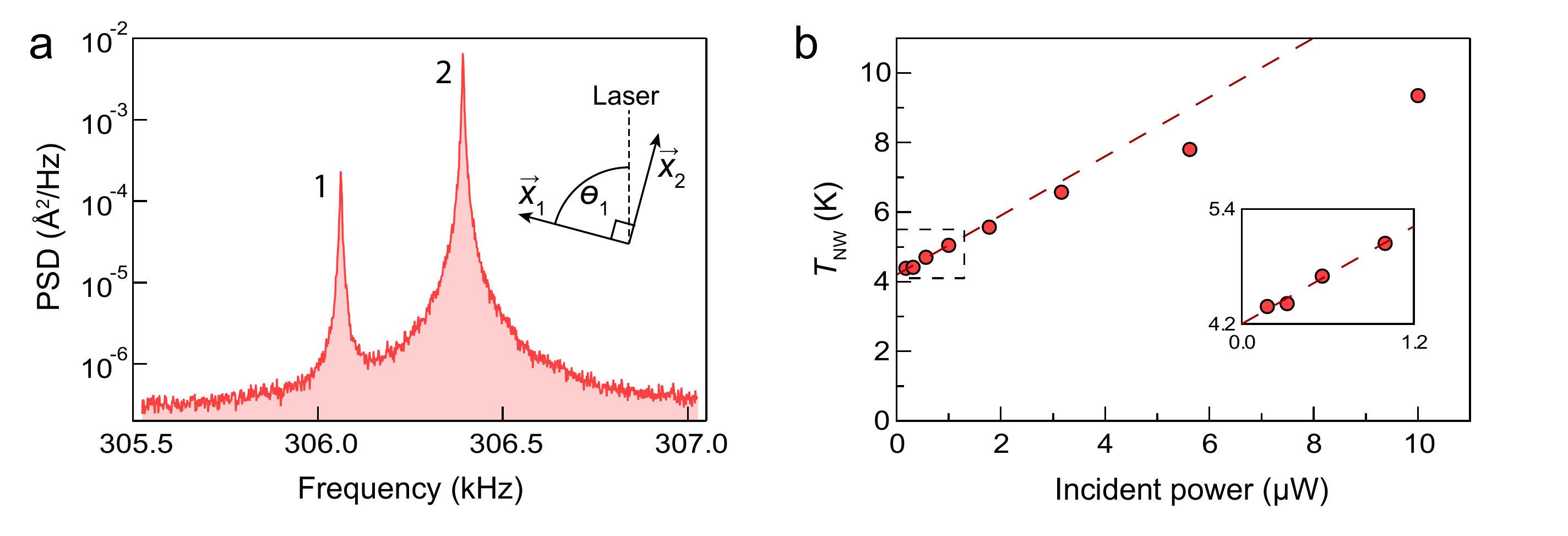}}
        \caption{\label{Fig_Tmeas} SiNW thermal displacement. (a) Representative power spectral density (PSD) of SiNW thermal oscillations at a base temperature of 4.2 K. Integrating this PSD gives $\left<x^2\right>$. Inset: diagram showing the approximate SiNW mode orientations relative to the optical axis.  (b) SiNW temperature vs. incident laser power data for a $\diameter 132$~nm SiNW. The dashed line shows a linear fit to the lowest three data points starting from 4.2~K at 0~$\upmu$W which is used to set the scale for the temperature axis. Inset: zoomed view of lowest four data points.}
\end{figure}

The equipartition theorem was used to determine SiNW temperature: 
\begin{align}\label{Eq_equi1}
\frac{1}{2}k_\text{B} T_{\text{NW}}=\frac{1}{2}k_n\left<x_n^2\right>,
\end{align}
where $k_\text{B}$ is the Boltzmann constant, $n$ represents a single flexural mode, $k_n$ is the spring constant of that mode, and $\left<x_n^2\right>$ is the time-averaged squared displacement of the $n$ mode. Since the lowest-frequency two modes are nearly degenerate and randomly oriented, the combined projection of both modes onto the optical axis was measured, given by: $\left<x^2\right>=\text{cos}^2(\theta_1)\left<x_1^2\right>+\text{sin}^2(\theta_1)\left<x_2^2\right>$. Assuming $k=k_1=k_2$, and using (\ref{Eq_equi1}), gives $k\left<x^2\right>=k_\text{B}T_{\text{NW}}$.
However, for this equation, $\left<x^2\right>$ must be measured at the tip of the nanowire. Due to the $2~\upmu$m spot size, it is difficult to locate the tip accurately, and one is forced to average over some length of the NW. Instead, $\left<x^2\right>$ was measured at multiple laser powers, including at several very low powers. A line was fit to the lowest three values of $\left<x^2\right>$. The temperature was assumed to be 4.2~K at zero laser power, and the $\left<x^2\right>$ values were scaled accordingly to produce Fig. \ref{Fig_Tmeas}b.

\begin{figure}[!htb]
        \center{\includegraphics[width=0.6\textwidth]
        {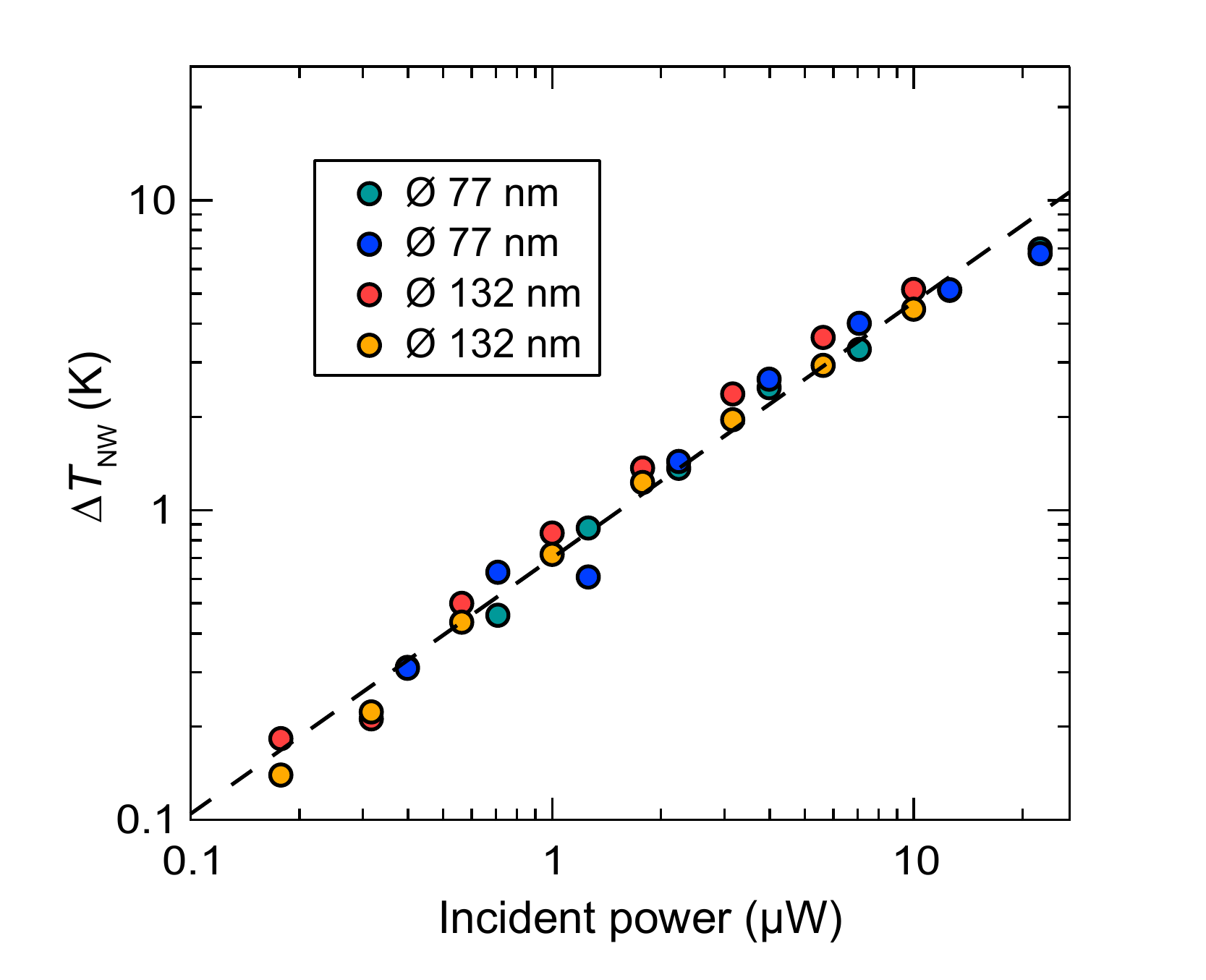}}
        \caption{\label{Fig_HvsP} Heating as a function of incident optical power for four nanowires, two from each array. The dashed line is the power law fit, $\Delta T_{\text{NW}}=0.70P^{0.83}$, to the combined data from all four SiNWs.}
\end{figure}

Heating and displacement sensitivity measurements were made on four SiNWs, two from each array, at 4.2 K. Heating was observed to be similar for all four SiNWs for a given incident power, as shown in Fig. \ref{Fig_HvsP}, meaning that it was independent of nanowire diameter. Although this was not studied in detail, the thermal resistance at a given temperature is approximately inversely proportional to the cross-sectional area, suggesting that the absorbed power is roughly directly proportional to the cross-sectional area of the SiNW. We fit the heating data with a power law, $\Delta T_{\text{NW}}=\text{A}P^\text{B}$, where P is the incident power, and A and B are the fit parameters. This relationship between incident power and SiNW heating was used to produce the heating axis in Fig.~3.

To extend measurements to lower temperatures but maintain displacement sensitivity around $1\times10^{-12}~\text{m/}\sqrt{\text{Hz}}$ one could attempt to modify the optical interactions with the nanowire, or one could improve the optics electronics. As seen in Fig.~3, the shot noise is significantly lower than the total noise, which was dominated by noise from the transimpedance amplifier. One change that could substantially improve the displacement sensitivity at low incident powers would be switching from a PIN photodiode to an avalanche photodiode. Photocurrent gains of 10 or more are possible with InGaAs avalanche photodiodes, meaning that one could obtain the same signal strength with 10 times less incident power, and therefore much less heating.


\bibliography{biblio}